\documentclass[a4paper,UKenglish]{lipics-v2016}

\usepackage{microtype}
\usepackage{amsmath}
\usepackage{amssymb}
\usepackage{amsthm}
\usepackage{amscd}
\usepackage{graphicx}
\usepackage{tikz,wrapfig}

\theoremstyle{definition}
\newtheorem{proposition}[theorem]{Proposition}

\newtheorem*{claim}{Claim}
\newtheorem*{proposition*}{Proposition}
\newcommand{\dom}{\operatorname{dom}}
\newcommand{\id}{\textnormal{id}}
\newcommand{\Cantor}{{\{0, 1\}^\mathbb{N}}}

\newcommand{\hide}[1]{}
\newcommand{\mto}{\rightrightarrows}
\newcommand{\uint}{{[0, 1]}}

\newcommand{\cc}{\textrm{CC}}
\newcommand{\UC}{\textrm{UC}}
\newcommand{\lpo}{\textrm{LPO}}
\newcommand{\llpo}{\textrm{LLPO}}
\newcommand{\aou}{\textrm{AoUC}}
\newcommand{\aouc}{\textrm{AoUC}_\uint}
\newcommand{\Ctwo}{C_{\{0, 1\}}}
\newcommand{\name}[1]{\textsc{#1}}
\newcommand{\C}{\textrm{C}}

\newcommand{\rDiv}{\operatorname{rDiv}}
\newcommand{\urDiv}{\operatorname{ubrDiv}}

\newcommand{\leqW}{\leq_{\textrm{W}}}
\newcommand{\nleqW}{\nleq_{\textrm{W}}}
\newcommand{\leW}{<_{\textrm{W}}}
\newcommand{\equivW}{\equiv_{\textrm{W}}}

\DeclareMathOperator{\arccot}{arccot}

\newcommand{\gauss}{\operatorname{GaussElim}}

\newcommand{\lrangle}[1]{\langle #1 \rangle}
\newcommand{\res}{\upharpoonright}
\newcommand{\fr}{^{\smallfrown}}

\newcommand{\longversion}[1]{#1}

\title{Dividing by zero -- how bad is it, really?\footnote{The first author was partially supported by a Grant-in-Aid for JSPS fellows. The second author was partially supported by the ERC inVEST (279499) project.}}
\titlerunning{Dividing by zero -- how bad is it, really?} 

\author[1]{Takayuki Kihara}
\author[2]{Arno Pauly}
\affil[1]{Department of Mathematics\\ University of California, Berkeley, United States\\
  \texttt{kihara@math.berkeley.edu}}
\affil[2]{D\'epartement d'Informatique\\ Universit\'e Libre de
Bruxelles, Belgium\\
  \texttt{Arno.M.Pauly@gmail.com}}
\authorrunning{T.~Kihara \& A.~Pauly} 

\Copyright{Takayuki Kihara and Arno Pauly}


\begin{document}

\maketitle

\begin{abstract}
In computable analysis testing a real number for being zero is a fundamental example of a non-computable task. This causes problems for division: We cannot ensure that the number we want to divide by is not zero. In many cases, any real number would be an acceptable outcome if the divisor is zero - but even this cannot be done in a computable way.

In this note we investigate the strength of the computational problem \emph{Robust division}: Given a pair of real numbers, the first not greater than the other, output their quotient if well-defined and any real number else. The formal framework is provided by Weihrauch reducibility. One particular result is that having later calls to the problem depending on the outcomes of earlier ones is strictly more powerful than performing all calls concurrently. However, having a nesting depths of two already provides the full power. This solves an open problem raised at a recent Dagstuhl meeting on Weihrauch reducibility.

As application for \emph{Robust division}, we show that it suffices to execute Gaussian elimination.
 \end{abstract}

\section{Introduction}

\emph{We cannot divide by zero!} is probably the first mathematical impossibility statement everyone encounters. In the setting we see it first, arithmetic of concrete integers, this does not cause any problems: Since it is obvious whether some number is zero or not, we simply refrain from attempting it -- and the multiplicative absorption of $0$ ensures that we have no reason for an attempt anyway. As our mathematical world expands to include more kinds of numbers, and in particular variables, we may have to introduce case distinctions at times in order to avoid this problem\footnote{Forgetting about these cases has probably caused a lot of anguish to pupils learning the outcome of their exams.}.

In most practical situations, this may seem unproblematic. However, a fundamental observation by \name{Brouwer} in the early development of constructive mathematics was that \emph{we cannot in general decide whether a real number is zero or not}. Thus, a case distinction based on whether our intended denominator is zero or not is not constructive. In a constructive setting, we can only divide by a number we know to be different from zero.

To consider a concrete example where we might want to divide by a number that could be zero, consider $a, b \in \mathbb{R}$ with $0 \leq a \leq b$, and the linear equation $a = bx$. We know that there is a solution $x_0 \in [0,1]$: If $b \neq 0$, then $x_0 := \frac{a}{b}$, otherwise $b = a = 0$, and any $x$ works. We see that we do not actually care about whether $b = 0$ or not, and we do not even need any particular outcome of a misguided attempt to calculate $\frac{0}{0}$ -- any number would do.

Unfortunately, the algorithm to divide a real number $a$ by a real number $b$ starts with searching for a rational number bounding $b$ away from $0$. If no such number exists, there will be no output at all, rather than some arbitrary number. The \emph{robust division} we would like to employ to solve linear equations as above is not actually computable.

In this note, we study the extent of non-computability of robust division in the formal setting of Weihrauch reducibility. Some results had already been obtained in \cite{paulyincomputabilitynashequilibria}. We will recall that robust division lies strictly in between the traditional non-constructive principles $\llpo$ and $\lpo$ and some other basic properties. Our concern then is with the question how multiple uses of robust division interact. We show that sequential uses of robust division cannot be reduced to parallel uses -- however, it suffices to have a nesting depths of $2$.

In \cite{paulyincomputabilitynashequilibria}, finding the solution to systems of linear inequalities via a modified Fourier-Motzkin elimination, and finding Nash equilibria in bimatrix games were explored as applications of robust division. Here, we shall consider Gaussian elimination as additional example.

\section{Background}

\subsection*{Computability on the reals and other represented spaces}
The long history of studying computability on the real numbers presumably goes back to \name{Borel} \cite{borel} (see \cite{brattka-history} for a detailed historical picture). Here, we follow the school of \name{Weihrauch} \cite{weihrauchd}. Computability is initially introduced over $\Cantor$ by means of Type-2 machines. These are obtained from the usual Turing machine model via a simple modification: The head on output tape can move to the right only (and in particular does so whenever a symbol is written), and the machines never halt. The restriction on the output tape ensures that as the computation proceeds, longer and longer finite prefixes of the ultimate infinite output are available.

The transfer of computability from $\Cantor$ to the spaces of actual interest is achieved via the notion of a represented space. For a more detailed introduction to the theory of represented spaces, we refer to \cite{pauly-synthetic}. A represented space is a pair $\mathbf{X} = (X, \delta_X)$ of a set $X$ and a partial surjection $\delta_X : \subseteq \Cantor \to X$ (the representation).

A multi-valued function\footnote{For a discussion of the notion of a multi-valued function, and in particular the difference to the notion of a relation, we refer to \cite{paulyziegler}, \cite{paulysearchproblems}.}  between represented spaces is a multi-valued function between the underlying sets. For $f : \subseteq \mathbf{X} \mto \mathbf{Y}$ and $F : \subseteq \Cantor \to \Cantor$, we call $F$ a realizer of $f$ (notation $F \vdash f$), iff $\delta_Y(F(p)) \in f(\delta_X(p))$ for all $p \in \dom(f\delta_X)$.
 $$\begin{CD}
\Cantor @>F>> \Cantor\\
@VV\delta_\mathbf{X}V @VV\delta_\mathbf{Y}V\\
\mathbf{X} @>f>> \mathbf{Y}
\end{CD}$$
A map between represented spaces is called computable (continuous), iff it has a computable (continuous) realizer. Note that a priori, the notion of continuity for maps between represented spaces differs from topological continuity. For the admissible represented spaces (in the sense of \cite{schroder}), the two notions do coincide, if a represented space is equipped with the final topology inherited from Cantor space along the representation. All representations we are concerned with in this note are admissible.

Before we introduce the standard representation of the real numbers, we fix some standard enumeration $\nu_\mathbb{Q} : \mathbb{N} \to \mathbb{Q}$ of the rationals. Now we define $\rho :\subseteq \Cantor \to \mathbb{R}$ via $\rho(0^{n_0}10^{n_1}1\ldots) = x$ iff $\forall i \in \mathbb{N} \ |\nu_\mathbb{Q}(n_i) - x| < 2^{-n}$. Note that using e.g.~the binary or decimal expansion would not have worked satisfactorily\footnote{As already noted by \name{Turing} \cite{turingb}.}. The choice of $\rho$ ensures that, informally spoken, every naturally encountered continuous function on the reals will be computable.

The naturals are represented in the obvious way by $\delta_\mathbb{N}(0^n1^\mathbb{N}) = n$. The finite spaces $\{0,\ldots,n\}$ are just the corresponding subspaces of $\mathbb{N}$. Likewise, we introduce the represented space $\uint$ as a subspace of $\mathbb{R}$.

For any represented space $\mathbf{X}$, there is a canonical definition of the represented space $\mathcal{A}(\mathbf{X})$ of closed subsets of $\mathbf{X}$. We only require this for the specific choices of $\mathbf{X} = \uint,\{0,\ldots,n\}$: In the former case, a closed subset is a closed subset in the usual sense, and it is represented by a list of rational open balls exhausting its complement in $\uint$. In the latter, any subset of $\{0,\ldots,n\}$ is an element of $\mathcal{A}(\{0,\ldots,n\})$, and a set $A$ is represented by $p \in \Cantor$ iff $01^k0$ occurs somewhere in $p$ iff $k \notin A$ for any $k \in \{0,\ldots,n\}$.

As there are canonical tupling functions $\langle \ldots \rangle : (\Cantor)^n \to \Cantor$ available, we can define products of represented spaces in a straight-forward way. We obtain binary and countable disjoint unions by $(\delta_0 + \delta_1)(0p) = \delta_0(p)$ and $(\delta_0 + \delta_1)(1p) = \delta_1(p)$, and $(\coprod_{i \in \mathbb{N}} \delta_i)(0^n1p) = \delta_n(p)$. We will iterate the binary product, starting with the convention $\mathbf{X}^0 = \{0\}$ and setting $\mathbf{X}^{n+1} = \mathbf{X}^n \times \mathbf{X}$. Finally, $\mathbf{X}^*$ is shorthand for $\coprod_{i \in \mathbb{N}} \mathbf{X}^i$.

\subsection*{Weihrauch reducibility}
Weihrauch reducibility is a computable many-one reduction comparing multi-valued functions between represented spaces. So $f \leqW g$ informally means that $f$ could be computed with the help of a single oracle-call to $g$.

		\begin{wrapfigure}{r}{3.5cm}
				\begin{tikzpicture}
					\node at (.5,4.75) {\small{name of some $y\in f(x)$}};
					\draw (-1.25,-.5) rectangle (2.25,4.25);
					\draw (0,0) rectangle (2,1);
					\node at (1,.5) {$H$};
					\draw[->] (0,-.5) -- (.25,0);
					\node at (.5,-1.25) {name of $x\in \dom(f)$};
					\draw[->] (.25,1) -- (.25,1.5);
					\node at (1.25,1.25) {\small name of $z$};
					\draw (0,1.5) rectangle (2,2.5);
					\node at (1,2) {$G$};
	                \node at (1.25,2.85) {\tiny{name of}};
					\node at (1.25,2.70) {\tiny{some $y \in g(z)$}};
					\draw[->] (.25,2.5) -- (.25,3);
					\draw[->] (0,-1) -- (0,-.5) -- (-.5,.5) -- (-.5,2) --(-.25,2.5) --(-.25,3);
					\draw (-1,3) rectangle (1,4);
					\node at (0,3.5) {$K$};
					\draw[->] (0,4) -- (0,4.5);
					\node at (1.6,3.4) {F};
				\end{tikzpicture}
				\vspace{-.75cm}
				\caption{Illustrating Definition \ref{def:weihrauch}\footnotemark }\label{fig:Wheirauch reduction}
				\vspace{-.75cm}
			\end{wrapfigure}
\footnotetext{This figure was taken from \cite{pauly-steinberg-csr}.}
			\begin{definition}\label{def:weihrauch}

				Let $f : \subseteq \mathbf{X} \mto \mathbf{Y}$ and $g : \subseteq \mathbf{U} \mto \mathbf{V}$ be partial multivalued functions between represented spaces.
				Say that $f$ is Weihrauch reducible to $g$, in symbols $f\leqW g$, if there are computable
				functions $K:\subseteq \Cantor\times \Cantor \to\Cantor$ and $H:\subseteq\Cantor\to\Cantor$ such that whenever $G$ is a realizer of $g$, the function $F := \left (p \mapsto K(p,G(H(p)))\right )$ is a realizer for $f$.
			\end{definition}

Based on earlier work by \name{Weihrauch} \cite{weihrauchb,weihrauchc}, Weihrauch reducibility was suggested as a framework for computable metamathematics in \cite{brattka2,brattka3} (see also \cite{gherardi,paulyincomputabilitynashequilibria}). We point to the introduction of \cite{hoelzl} for a recent overview on the development of the field so far.

We shall denote the set of Weihrauch degrees by $\mathfrak{W}$, and point out some operations on them. As shown in \cite{paulyreducibilitylattice}, the binary product $\times$, the binary disjoint union $\sqcup$, the countable disjoint union $\coprod$ and the operation $^*$ all can be lifted from represented spaces via multivalued functions between represented spaces to Weihrauch degrees. $\mathfrak{W}$ is a distributive lattice, and $\sqcup$ is the join. However, no non-trivial countable suprema exist in $\mathfrak{W}$ as shown in \cite{paulykojiro}. In particular, $\coprod$ is not the countable join.

Informally, $f \sqcup g$ means that both $f$ and $g$ are available for use, but the user has to decide for each instance on one of the two to call. A call to $f \times g$ means making two independent calls, one to $f$ and one to $g$. Using $f^*$ means that we first decide on some number $n \in \mathbb{N}$, and then make $n$ independent calls to $f$.

We want to use a further operation; corresponding to first making a call to some $g$ and then a call to $f$ depending on the outcome of the call to $g$. In \cite{paulybrattka3cie,gherardi4} the operation $\star$ was defined as $f \star g := \max_{\leqW} \{f' \circ g' \mid f' \leqW f \wedge g' \leqW g\}$. Here the maximum is understood to range over all $f'$, $g'$ with types such that $f' \circ g'$ is well-defined. While it is not obvious that this maximum exists, an explicit construction is provided in \cite{paulybrattka4}. Informally, an input to $f \star g$ consists of an input to $g$ together with a multivalued function computing some input to $f$ from an output to $g$. The output is the output of $g$ together with the output of $f$.

We iterate both $\times$ and $\star$: $f^0 = f^{(0)} = \id_\Cantor$, and $f^{n+1} = f^n \times f$ and $f^{(n+1)} = f^{(n)} \star f$.

\subsection*{Special Weihrauch degrees}
We will refer to a number of well-studied specific Weihrauch degrees in this paper. We shall first recall the degrees $\lpo$ and $\llpo$ from \cite{weihrauchc}, the Weihrauch degree counterparts to the Brouwerian counterexamples in intuitionistic mathematics. $\lpo : \Cantor \to \{0,1\}$ maps $0^\mathbb{N}$ to $1$ and each $p \neq 0^\mathbb{N}$ to $0$. This map is equivalent to the characteristic function of $0$ in $\mathbb{R}$ or $\uint$, and we will not distinguish these maps. The map $\llpo : \Cantor \mto \{0,1\}$ outputs $0$ if the first $1$ in the input occurs at an even position, and $1$ if it occurs at an odd position. An alternative characterization would be $\llpo : \mathbb{R} \times \mathbb{R} \to \{0,1\}$ with $0 \in \llpo(0,y)$, $1 \in \llpo(x,0)$ and $0,1 \in \llpo(x,y)$ if $x \neq 0 \neq y$. If the input was $0^\mathbb{N}$, both $0$ and $1$ are valid outputs. It holds that $\llpo \leW \lpo$.

A bountiful source of calibrating principles is found in the closed choice principles and their restrictions:
\begin{definition}
$\C_\mathbf{X} : \subseteq \mathcal{A}(\mathbf{X}) \mto \mathbf{X}$, $\dom(\C_\mathbf{X}) = \{A \in \mathcal{A}(\mathbf{X}) \mid A \neq \emptyset\}$, $x \in \C_\mathbf{X}(A) \Leftrightarrow x \in A$.
\end{definition}

Thus, closed choice is the task to find a point in a given non-empty closed set. As being non-empty is merely promised rather than constructively witnessed, this task is generally not computable: As long as there is more than one remaining choice, whenever we start outputting some point we might learn next that this point is not in the closed set after all (e.g.~by reading some rational ball containing it in the name of the closed set).

These principles have been extensively studied \cite{brattka3,paulybrattka,paulybrattka2,paulybrattka3cie,paulyleroux,hoelzl}. Depending on the topological properties of the space $\mathbf{X}$ and potentially restrictions to certain subsets, these principles have been revealed to be useful in characterizing many other principles.

Most relevant for us are the principles $\C_{\{0,\ldots,n\}}$. It was shown in \cite{paulyincomputabilitynashequilibria} that $\C_{\{0,\ldots,n\}} \leqW \C_{\{0,1\}}^n$, and it follows from the independent choice theorem in \cite{paulybrattka} that $\C_{\{0,1\}}^n \leqW \C_{\{0,1\}}^{(n)} \leqW \C_{\{0,\ldots,2^n-1\}}$. It is quite easy to see that $\C_{\{0,1\}} \equivW \llpo$.

We make passing references to $\C_\mathbb{N}$ (and use that $\C_\mathbb{N} \equivW \C_\mathbb{N} \star \C_\mathbb{N}$ by the independent choice theorem from \cite{paulybrattka}), to $\cc_\uint$, the restriction of $\C_\uint$ to connected subsets and to $\textrm{PCC}_\uint$, the restriction of $\C_\uint$ to connected sets with positive Lebesgue measure. We also mention $\star\textrm{-WWKL}$ from \cite{hoelzl}, which is $\coprod_{n \in \mathbb{N}} (2^{-n})\textrm{-WWKL}$, where $\varepsilon\textrm{-WWKL}$ is the restriction of $\C_\Cantor$ to sets with measure at least $\varepsilon$. By $\UC_\mathbf{X}$ we denote the restriction of $\C_\mathbf{X}$ to singletons \cite{paulybrattka}. Finally, $\C_{\sharp = 2}$ and $\C_{\sharp \leq 2}$ from \cite{paulyleroux} are the restrictions of $\C_\Cantor$ to sets with cardinality $2$ and at most $2$ respectively.
\section{Robust division}
We consider two variants of robust division: In one case, we know an upper bound on the result, in the second, we do not. Modulo the rescaling, the first case corresponds to knowing that the denominator is at least as big as the numerator.

\begin{definition}[\cite{paulyincomputabilitynashequilibria}]
\label{bimatrix:def:rdiv}
Define $\rDiv : \mathbb{R} \times \mathbb{R} \mto \uint$ via $\frac{\min \{|x|, |y|\}}{|y|} \in \rDiv(x, y)$ iff $y \neq 0$ and $z \in \rDiv(x, 0)$ for all $x \in \mathbb{R}, z \in \uint$.
\end{definition}

To simplify notation, we will usually assume that inputs $(x, y)$ for $\rDiv$ already satisfy $0 \leq x \leq y$, so that $\frac{\min \{|x|, |y|\}}{|y|} = \frac{x}{y}$ holds.

\begin{definition}
Define $\urDiv : \mathbb{R} \times \mathbb{R} \mto \mathbb{R}$ by $\frac{x}{y} \in \urDiv(x,y)$ iff $y \neq 0$ and $z \in \operatorname{ubrDiv}(x, 0)$ for all $x, z \in \mathbb{R}$.
\end{definition}

It turns out that the case distinction on $y \neq 0$ or $y = 0$ is equivalent to the unbounded case $\urDiv$. Thus, we do not need to investigate $\urDiv$ as an independent basic operation. Note that the following proof also establishes that it makes no difference for the degree of $\urDiv$ if the result is presumed to be non-negative.

\begin{proposition}
$\urDiv \equivW \lpo$
\begin{proof}
The direction $\urDiv \leqW \lpo$ follows from computability of division where well-defined and the definition.

For the other direction, note that given some $p \in \Cantor$ we can compute $x, y \in \mathbb{R}$ such that if $p \neq 0^\mathbb{N}$, then $\frac{x}{y} = \min\{n \in \mathbb{N} \mid p(n) = 1\}$ and $x = y = 0$ if $p = 0^\mathbb{N}$. Furthermore, there is a computable multivalued retract $\rho : \mathbb{R} \mto \mathbb{N}$, so we may pretend that the output of $\operatorname{ubrDiv}(x,y)$ is a natural number $n$ indicating the position of the first $1$ in $p$, if it exists. Given this number, we can then check whether $p(n)$ is $1$ or not, which in turn determines the answer to $\lpo(p)$.
\end{proof}
\end{proposition}

The bounded variant of robust division was already established as a \emph{new} degree in \cite{paulyincomputabilitynashequilibria}. We recall some results on this degree from the literature before continuing its investigation.

\begin{proposition}[\cite{paulyincomputabilitynashequilibria}]
\label{prop:rdivprior}
\begin{enumerate}
\item $\C_{\{0,1\}} \leW \rDiv \leW \lpo$.
\item $\rDiv \leW \cc_{\uint}$.
\item $\rDiv \nleqW \C_{\{0,1\}}^*$.
\end{enumerate}
\end{proposition}

\begin{proposition}[{\cite[Theorem 16.3, Corollary 16.5 \& Theorem 16.6]{hoelzl}}]
\begin{enumerate}
\item $\rDiv \leW \textrm{PCC}_\uint$.
\item $\rDiv$ is join-irreducible.
\item $\rDiv \nleqW *\textrm{-WWKL}$.
\end{enumerate}
\end{proposition}

The preceding results from \cite{hoelzl} intuitively state that there is a mechanism to solve $\rDiv$ in a probabilistic way with positive probability and error detection. However, there is no way to obtain a positive lower bound on the probability of solving a given instance correctly.

It is a well-known phenomenon in the study of Weihrauch reducibility that closed choice principles make very convenient representatives of Weihrauch degrees (cf.~\cite{brattka3,paulybrattka,paulybrattka2,paulybrattka3cie,paulyleroux}). The case of robust division is no different: For a represented space $\mathbf{X}$ we denote by $\aou_\mathbf{X}$ the restriction of $\C_\mathbf{X}$ to $\{A \in \mathcal{A}(\mathbf{X}) \mid |A| = 1\} \cup \{X\}$ following an idea of \name{Brattka}. Just by its definition, it is clear that $\textrm{UC}_\mathbf{X} \leq_W \aou_\mathbf{X} \leq_W \C_\mathbf{X}$ holds for any space $\mathbf{X}$. In the following we shall focus on $\aou_\uint$.

\begin{proposition}\footnote{This result was suggested to the author by \name{Brattka}, and has been shown in \cite{paulyphd}.}
\label{bimatrix:prop:aouequivrdiv}
$\rDiv \equivW \aou_\uint$
\begin{proof}
The reduction $\rDiv \leqW \aou_\uint$ is straight-forward: On input $(x, y) \in \mathbb{R}^2$ for $\rDiv$, while the search for a $k \in \mathbb{N}$ with $y > 2^{-k}$ continues, the input to $\aou_\uint$ is kept at $\uint$. If such a $k$ is ever found, one can compute $\frac{x}{y}$, and hence also $\{\frac{x}{y}\}$ as $\uint$ is computably Hausdorff and collapse the unit interval to it.

For the other direction, as long as the input to $\aou_\uint$ has not collapsed, one starts to input $(0, 0)$ to $\rDiv$. If the input of $\aou_\uint$ ever collapsed to $\{z\}$, one can compute $z$. The input to $\rDiv$ can still be chosen from some interval $[0, 2^{-k}] \times [0, 2^{-k}]$. In particular, $x = 2^{-k}z$ and $y = 2^{-k}$ works and forces the correct output.
\end{proof}
\end{proposition}

\longversion{We shall now consider $\aou_\mathbf{X}$ for some further choices of a space $\mathbf{X}$. Essentially, the degree seems to depend primarily on compactness of $\mathbf{X}$.
\begin{proposition}
$\aou_\mathbb{R} \equiv_W \aou_\mathbb{N} \equiv_W \C_\mathbb{N}$
\begin{proof}
One direction follows from $\C_\mathbb{N} \equiv_W \textrm{UC}_\mathbb{N} \equiv_W \textrm{UC}_\mathbb{R} \leq_W \aou_\mathbb{R}$, where each reduction is either trivial or was shown in \cite{paulybrattka}. For the other direction, note that the distinction between $|A| = 1$ and $A = \mathbb{R}$ for some $A \in \mathcal{A}(\mathbb{R})$ is equivalent to $\lpo$. We thus obtain $\aou_\mathbb{R} \leq_W \textrm{UC}_\mathbb{R} \star \lpo \leq_W \C_\mathbb{N} \star \C_\mathbb{N} \equiv_W \C_\mathbb{N}$.
\end{proof}
\end{proposition}

\begin{corollary}
$\urDiv \leW \aou_\mathbb{R}$
\end{corollary}

\begin{proposition}
\label{prop:aoucompact}
For a computably compact computable Polish space $\mathbf{X}$ we have $\aou_\mathbf{X} \leqW \aou_\Cantor$.
\begin{proof}
The proof shares many ideas with \cite[Proposition 1.9]{paulyleroux}. Using the computable compactness of $\mathbf{X}$, we can effectively find a finitely branching tree labeled with open rational balls, such that the balls used on the $n$-th level have radius $2^{-n}$, and the ball of any vertex is covered by the union the balls of its children. We shall assume that $\mathbf{X}$ itself has radius $1$, and is used as the label of the root. Note that the centers of the balls along any path through the tree form a fast Cauchy sequence, and recall that limits of fast Cauchy sequences are computable in computable Polish spaces.

Given some $A \in \mathcal{A}(\mathbf{X})$ for computably compact $\mathbf{X}$, we can recognize $A \subseteq B$ for open balls $B$. In particular, we will use $A \in \mathcal{A}(\mathbf{X})$ to determine a subtree of the tree introduced above by removing all siblings of a vertex as soon as we learn that its label covers $A$. If $A$ is a singleton $\{x\}$, the resulting tree will have a unique infinite branch, and we can recover $x$ from such a branch. If $A = \mathbf{X}$, no vertex is ever removed from the tree.

There is a standard correspondence between the finitely branching trees and the closed subsets of Cantor space which maps the trees constructed in the previous paragraph to valid input $\aou_\Cantor$, in a way that the output determines an infinite path through the tree. This in turn provides the answer to the original instance of $\aou_\mathbf{X}$.
\end{proof}
\end{proposition}

\begin{corollary}
$\aou_\uint \equivW \aou_{\uint^n} \equivW \aou_\Cantor$
\begin{proof}
One direction was given in Proposition \ref{prop:aoucompact}. For the other direction, note that any $\uint^n$ contains some closed subset $C$ homeomorphic to $\Cantor$, and furthermore admitting a computable multivalued retract $R_C : \uint^n \mto \Cantor$.
\end{proof}
\end{corollary}

We shall consider another representative of the Weihrauch degree of $\rDiv$. Define $\operatorname{DependentCut} : \Cantor \mto \Cantor$ via $\operatorname{DependentCut}(0^n1p) = \{p\}$ and $\operatorname{DependentCut}(0^\mathbb{N}) = \Cantor$.

\begin{proposition}
$\operatorname{DependentCut} \equiv_W \aou_\Cantor$
\begin{proof}
For the $\operatorname{DependentCut} \leq_W \aou_\Cantor$ direction, start by giving the full binary tree as input to $\aou_\Cantor$. As soon as a $1$ is found in the input to $\operatorname{DependentCut}$, i.e.~the input is to be determined as $0^n1p$, prune any branch of the tree not compatible with $p$ at the current level, and extend the tree to have $p$ as its unique path.

For the other direction, start writing $0$s. As soon as any branch in the input tree to $\aou_\Cantor$ is pruned, write a $1$. Now the tree is guaranteed to have a unique infinite path, and compactness and admissibility of $\Cantor$ mean that this path can be computed. We just do so, and output it after the $1$.
\end{proof}
\end{proposition}
}
\section{Sequential versus concurrent uses of $\rDiv$}
If multiple uses of some noncomputable principle are needed to solve a particular task, an important distinction is whether these have to be sequential, or can be applied in a concurrent fashion. In the former case, some instances to the principle may depend on outputs obtained from prior invocations. In the latter, each instantiation is independent of the others. For various principles, however, we find that sequential uses can be reduced to concurrent uses.

\begin{definition}
We call $f$ \emph{finitely concurrent}, iff $f^* \equivW \coprod_{n \in \mathbb{N}} f^{(n)}$.
\end{definition}

\begin{proposition}[\cite{paulyleroux}]
The following are finitely concurrent:
\begin{enumerate}
\item $\lpo$
\item $\llpo \equivW \Ctwo$
\item $\C_{\sharp = 2}$
\item $\C_{\sharp \leq 2}$
\end{enumerate}
\end{proposition}

Whether $\rDiv$ is finitely concurrent in this sense was posed as an open question during the Dagstuhl workshop \emph{Measuring the Complexity of Computational Content} in September 2015 \cite{pauly-dagstuhl}. We can now provide a negative answer\footnote{Which also disproves a claim made in the PhD thesis of the second author \cite[Theorem 5.2.1.6]{paulyphd}.}:

\begin{theorem}
\label{theo:main}
$\llpo \star \aouc \nleqW \aouc^k$ for all $k \in \mathbb{N}$.

\begin{proof}
We say that a binary tree $T\subseteq\{0,1\}^\ast$ is an {\em a.o.u.~tree} if for any height $n\in\mathbb{N}$ either $|T\cap\{0,1\}^n|=2^n$ or $|T\cap\{0,1\}^n|=1$.
Clearly, one can identify ${\rm AoUC}_{[0,1]}$ with the partial multi-valued function sending an a.o.u.~tree $T$ to all infinite paths through $T$.
We often identify a set $S\subseteq\{0,1\}^\ast$ with its characteristic function $\chi_S:\{0,1\}^\ast\to\{0,1\}$.
Under this identification, a partial function $t:\subseteq\{0,1\}^\ast\to\{0,1\}$ is called a {\em partial tree} if ${\rm Tr}(t):=\{\sigma\in{\rm dom}(t):t(\sigma)=1\}$ forms a subtree of $\{0,1\}^{\ast}$.
If such $t$ is computable, we call $t$ a {\em partial computable tree}.
Note that a tree is computable if and only if it is of the form ${\rm Tr}(t)$ for a partial computable tree $t$ which is {\em total}, that is, ${\rm dom}(t)=\{0,1\}^\ast$.
We say that a partial computable tree $t$ {\em looks like an a.o.u.~tree at $(l,s)$} if
\begin{enumerate}
\item $t(\sigma)[s]$ converges for any binary string $\sigma$ of length $l$,
\item for any $n<l$, the cardinality of ${\rm Tr}(t)\cap\{0,1\}^n$ is either $2^n$ or $1$,
\end{enumerate}
where $t(\sigma)[s]$ is the result of the computation of $t(\sigma)$ by stage $s$.
Note that given $(l,s)\in\mathbb{N}^2$, we can effectively decide whether $t$ looks like an a.o.u.~tree at $(l,s)$ or not.
By ${\rm aou}(t,s)$ we denote the greatest $l\leq s$ such that $t$ looks like an a.o.u.~tree at $(l,s)$.

Consider two partial multi-valued functions $Z_0={\rm AoUC}_{[0,1]}\times{\rm id}_\mathbf{X}$ and $Z_1=({\rm id}_{\{0,1\}^\mathbb{N}}\circ\pi_0,{\rm C}_{\{0,1\}}\circ{\rm eval})$, where $\mathbf{X}$ is the represented space $C(\{0,1\}^\mathbb{N},{\rm dom}(\mathrm{C}_{\{0,1\}}))$ of continuous functions from Cantor space $\{0,1\}^\mathbb{N}$ into the hyperspace ${\rm dom}(\mathrm{C}_{\{0,1\}})=\mathcal{A}(\{0,1\})\setminus\{\emptyset\}$ of nonempty closed subsets of $\{0,1\}$.
More explicitly, we consider the following two partial multi-valued functions:
\begin{align*}
Z_0&:{\rm dom}({\rm AoUC}_{[0,1]})\times \mathbf{X}\rightrightarrows\{0,1\}^\mathbb{N}\times \mathbf{X},\\
Z_0&(T,S)={\rm AoUC}_{[0,1]}(T)\times\{S\},\\
Z_1&:\{0,1\}^\mathbb{N}\times \mathbf{X}\rightrightarrows \{0,1\}^\mathbb{N}\times 2,\\
Z_1&(x,S)=\{x\}\times{\rm C}_{\{0,1\}}(S(x)).
\end{align*}

Clearly, $Z_0\leq_W{\rm AoUC}_{[0,1]}$ and $Z_1\leq_W{\rm C}_{\{0,1\}}$.
We will show that $Z_1\circ Z_0\not\leq_{\rm W}{\rm AoUC}_{[0,1]}^k$.
Let $\{(\mathbf{t}^e,\varphi_e,\psi_e)\}_{e\in\mathbb{N}}$ be an effective enumeration of all triples of $k$-tuples $\mathbf{t}^e=(t^e_i)_{i<k}$ of partial computable trees, partial computable functions $\varphi_e:\subseteq(\{0,1\}^\mathbb{N})^k\to\{0,1\}^\mathbb{N}$ and $\psi_e:\subseteq (\{0,1\}^\mathbb{N})^k\to \{0,1\}$.
Intuitively, $(\mathbf{t}^e,\varphi_e,\psi_e)$ is a triple constructed by the opponent ${\sf Opp}$, who tries to show $Z_1\circ Z_0\leq_{\rm W}{\rm AoUC}_{[0,1]}^k$ for some $k$.
The game proceeds as follows:
The proponent {\sf Pro} of our claim gives an instance $(T_r,S_r)$ of $Z_1\circ Z_0$, so that $(T_r,S_r)\in{\rm dom}({\rm AoUC}_{[0,1]})\times\mathbf{X}$.
In particular, $T_r$ is an a.o.u.~tree, and $S_r$ is a continuous function from $[T_r]$ into $\mathcal{A}(\{0,1\})$.
Then, ${\sf Opp}$ reacts with an instance $\mathbf{t}^r$ of ${\rm AoUC}_{[0,1]}^k$, that is, a $k$-tuple $\mathbf{t}^r=(t^r_i)_{i<k}$ of {\em total} a.o.u.~trees.
If {\sf Opp} wins, {\sf Opp} has to ensure that if $(p_i)_{i<k}$ is a $k$-tuple of infinite paths through {\sf Opp}'s a.o.u.~trees, that is, $p_i\in[{\rm Tr}(t^r_i)]$, then $\varphi_r((p_i)_{i<k})=x$ is a path through {\sf Pro}'s a.o.u.~tree $T_r$ and $\psi_r((p_i)_{i<k})$ chooses an element of {\sf Pro}'s set $S_r(x)$, where {\sf Opp} can use information on (names of) $T_r$ and $S_r$ to construct $\varphi_r$ and $\psi_r$.
Our purpose is to prevent ${\sf Opp}$'s strategy.

Given $e$, we will introduce the $e$-th strategy, which works as a proponent ${\sf Pro}$ of our claim.
The $e$-th strategy {\sf Pro} will construct a computable a.o.u.~tree $T_e\subseteq\{0,1\}^{*}$ and a computable function $S_e:\{0,1\}^\mathbb{N}\to\mathcal{A}(\{0,1\})$ in a computable way uniformly in $e$.
These will prevent ${\sf Opp}$'s strategy, that is, there is a $k$-tuple $(p_i)_{i<k}$ of infinite paths through {\sf Opp}'s a.o.u.~trees such that if $\varphi_e((p_i)_{i<k})=x$ chooses a path through {\sf Pro}'s a.o.u.~tree $T_e$ then $\psi_e((p_i)_{i<k})$ cannot be an element of {\sf Pro}'s set $S_e(x)$.

We will also define ${\tt state}(e,s)\in\{0,\dots,k\}\cup\{\tt end\}$.
The value ${\tt state}(e,s)=q$ for $q\not={\tt end}$ indicates that the $e$-th strategy ${\sf Pro}$ believes that by stage $s$, {\em at least $q$ many trees $(t_{u(j)}^e)_{j<q}$ in {\sf Opp}'s $k$-tuple $(t^e_i)_{i<k}$ have been forced not to have more than one infinite path}, that is, ${\rm Tr}(t^e_{u(j)})$ for each $j<q$ has a unique infinite path whenever the opponent ${\sf Opp}$ has a chance of winning this game with the triple $(\mathbf{t}^e,\varphi_e,\psi_e)$.
Under this assumption, if ${\tt state}(e,s)=q$, the fact that these $q$ many trees has no more than one infinite path will be witnessed at some stage.
The value ${\tt state}(e,s)={\tt end}$ indicates that the winning of {\sf Pro} is already witnessed by {\sf Pro}'s action of shrinking {\sf Pro}'s a.o.u.~tree $T_e$ to a tree having a unique path which avoids all $\varphi_e$-values made by {\sf Opp}.

By induction on $s$, we determine the set $T_e\cap\{0,1\}^s$ of strings in $T_e$ of length $s$ and ${\tt state}(e,s)$.
In the beginning of our construction, we define ${\tt state}(e,0)=0$.
At stage $s$, we inductively assume that $T_e\cap\{0,1\}^{s-1}$ and ${\tt state}(e,s-1)$ have already been defined, say ${\tt state}(e,s-1)=q$, and that if ${\tt state}(e,s-1)\not={\tt end}$ then $T_e\cap\{0,1\}^{s-1}=\{0,1\}^{s-1}$.

At stage $s$, the $e$-th strategy {\sf Pro} acts as follows:

\begin{enumerate}
\item Ask whether there exists $l\leq s$ such that $t^e_i$ for each $i<k$ looks like an a.o.u.~tree at $(l,s)$, and at least $q$-many trees among {\sf Opp}'s $k$-tuple $(t^e_i)_{i<k}$ have no more than one node above height $l$.
In other words, for ${\rm aou}(e,s):=\min_{i<k}{\rm aou}(t^e_i,s)$, ask whether there are at least $q$ many $i<k$ such that $|{\rm Tr}(t^e_i)\cap \{0,1\}^{{\rm aou}(e,s)}|=1$.
\begin{enumerate}
\item If no, we go to the next stage $s+1$ after setting ${\tt state}(e,s)={\tt state}(e,s-1)$ and $T_e\cap\{0,1\}^{s}=\{0,1\}^s$.
\item If yes, go to item (2).
\end{enumerate}
\item Ask whether $\varphi_e((p_i)_{i<k})$ already computes a node (of {\sf Pro}'s a.o.u.~tree $T_e$) of length at least $q+1$ for any $k$-tuple of paths $p_i$ through {\sf Opp}'s a.o.u.~trees $t^e_i$, that is,
\begin{align*}
(\forall i<k)(\forall(\sigma_i)_{i<k})\;[&((\forall i<k)\;\sigma_i\in {\rm Tr}(t^e_i)\cap \{0,1\}^{{\rm aou}(e,s)})\\
&\rightarrow\;(\forall m\leq q)\;\varphi_e((\sigma_i)_{i<k})(m)[s]\downarrow].
\end{align*}

Here, by effective continuity, the Type-2 computation $\varphi_e:(\{0,1\}^\mathbb{N})^k\to\{0,1\}^\mathbb{N}$ is approximated by a Type-1 computation $\tilde{\varphi}_e:(\{0,1\}^\ast)^k\to\{0,1\}^\ast$ (e.g., consider the Type-2 Turing machine model).
We always identify $\varphi_e$ with $\tilde{\varphi}_e$, and therefore, the notation $\varphi_e((\sigma_i)_{i<k})(m)$ makes sense, that is, by $\varphi_e((\sigma_i)_{i<k})(m)[s]\downarrow$, we mean that the computation of $\tilde{\varphi}_e((\sigma_i)_{i<k})(m)$ halts by stage $s$.
\begin{enumerate}
\item If no, we go to the next stage $s+1$ after setting ${\tt state}(e,s)={\tt state}(e,s-1)$ and $T_e\cap\{0,1\}^{s}=\{0,1\}^s$.
\item If yes, go to item (3).
\end{enumerate}
\item Ask whether the image of the product of $k$ many closed sets generated by {\sf Opp}'s $k$-tuple $\mathbf{t}^e$ under the map $\varphi_e$ covers the whole space $\{0,1\}^\mathbb{N}$.
Formally speaking, let us consider the following set:
\[\varphi_e[{\rm Tr}(\mathbf{t}^e)\cap \{0,1\}^l]\res p:=\{\varphi_e((\sigma_i)_{i<k})\res p:(\forall i<k)\;\sigma_i\in{\rm Tr}({t}^e_i)\cap \{0,1\}^l\},\]
and then ask whether $\tau\in\varphi_e[{\rm Tr}(\mathbf{t}^e)\cap \{0,1\}^{{\rm aou}(e,s)}]\res q+1$ for all $\tau\in \{0,1\}^{q+1}$.
\begin{enumerate}
\item If no, choose a witness $\tau$, and we finish the construction by setting ${\tt state}(e,s)={\tt end}$ after defining $T_e$ as a tree having a unique infinite path $\tau\fr 0^\infty:=\tau\fr 000\dots$.
\item If yes, go to item (4).
\end{enumerate}
\item  Ask whether $\psi_e((p_i)_{i<k})$ already computes some value $j\in\{0,1\}$ for any $k$-tuple of paths $p_i$ through {\sf Opp}'s a.o.u.~trees $t^e_i$, that is,
\[(\forall i<k)(\forall(\sigma_i)_{i<k})\;[((\forall i<k)\;\sigma_i\in {\rm Tr}(t^e_i)\cap\{0,1\}^{{\rm aou}(e,s)})\;\rightarrow\;\psi_e(\mathbf{\sigma})[s]\downarrow].\]
\begin{enumerate}
\item If no, go to the next stage $s+1$ after setting ${\tt state}(e,s)={\tt state}(e,s-1)$ and $T_e\cap \{0,1\}^{s}=\{0,1\}^s$.
\item If yes, let $D_{e,s}=\{\mathbf{\sigma}\in{\rm Tr}(\mathbf{t}^e)\cap \{0,1\}^{{\rm aou}(e,s)}:\varphi_e(\mathbf{\sigma})\succeq 0^q1\}$.
Note that $D_{e,s}\not=\emptyset$ since we answered yes in item (3); therefore $\varphi_e[{\rm Tr}(\mathbf{t}^e)\cap \{0,1\}^{{\rm aou}(e,s)}]\res q+1=\{0,1\}^{q+1}$.
If $i\not\in\psi_e[D_{e,s}]$ for some $i\in\{0,1\}$, then remove $1-i$ from $S_e(0^q1)$ (hence, we have $S_e(0^q1)=\{i\}$).
If $\psi_e[D_{e,s}]=\{0,1\}$ then remove $0$ from $S_e(0^q1)$.
In these cases, set ${\tt state}(e,s)=q+1$ and $T_e\cap\{0,1\}^{s}=\{0,1\}^s$.
\end{enumerate}
\end{enumerate}

Eventually, $T_e$ is constructed as an a.o.u.~tree, and $S_e(x)\in{\rm dom}({\sf C}_2)$.

\begin{claim}
Assume that $(t^e_i)_{i<k}$ determines a $k$-tuple of a.o.u.~trees.
Then, there is a realizer $G$ of ${\rm AoUC}_{[0,1]}^k$ such that $(\varphi_e\circ G((t^e_i)_{i<k}),\psi_e\circ G((t^e_i)_{i<k}))$ is not a solution to $Z_1\circ Z_0(T_e,S_e)$, that is, $\varphi_e\circ G((t^e_i)_{i<k})\not\in[T_e]$ or otherwise $\psi_e\circ G((t^e_i)_{i<k})\not\in S_e\circ\varphi_e\circ G((t^e_i)_{i<k})$, where $[T_e]$ denotes the set of all infinite paths through $T_e$.
\end{claim}

\begin{proof}
Assume that $\mathbf{t}^e=(t^e_i)_{i<k}$ determines a $k$-tuple of a.o.u.~trees.
In this case, $t^e_i$ is a total tree for each $i<k$.
Suppose for the sake of contradiction that the conclusion fails (that is, {\sf Opp} wins).
Then $\varphi_e$ and $\psi_e$ are defined on all tuples of infinite paths through ${\rm Tr}(t^e_i)$, $i<k$.
Since $q=0$ at first, the condition in item (1) is automatically fulfilled.
Note that since $t^e_i$ is a total a.o.u.~tree, the value ${\rm aou}(e,s)$ tends to infinity as $s\to\infty$.
Therefore, since $\varphi_e$ is defined on all paths of {\sf Opp}'s trees, by compactness, the condition in item (2) is also satisfied at some stage $s$.
If $\tau\not\in\varphi_e[{\rm Tr}(\mathbf{t}^e)\cap \{0,1\}^{{\rm aou}(e,s)}]\res q+1$, then $T_e$ has a unique infinite path $\tau\fr 0^\infty$; therefore $\varphi_e\circ G(\mathbf{t}^e)\not\in[T_e]$ for any realizer $G$, which contradicts our assumption.
Therefore, $\varphi_e[{\rm Tr}(\mathbf{t}^e)\cap\{0,1\}^{{\rm aou}(e,s)}]\res q+1=\{0,1\}^{q+1}$.
By compactness, the condition in item (4) is eventually satisfied.
In any cases, for some $\mathbf{\sigma}=(\sigma_i)_{i<k}\in D_{e,s}$, $\psi_e(\mathbf{\sigma})\not\in S_e(\varphi_e(\mathbf{\sigma}))$ by our construction.
In order for {\sf Opp} to win this game, {\sf Opp} has to declare that $\sigma_i$ for some $i<k$ is not extendible to an infinite path through $t^e_i$.
Consequently, under our assumption that {\sf Opp} wins, {\sf Pro}'s strategy forces such $t^e_i$ not to be the full binary tree; therefore $t^e_i$ has only one path since $t^e_i$ is an a.o.u.~tree.
Then we continue the same argument with $q=1$.
We can still satisfy the condition in item (1) at some stage since we know at most one tree $t^e_i$ has only one path.
Eventually, this construction forces that any of $t^e_i$ has only one path.
Then, however, it is impossible to satisfy $\varphi_e[{\rm Tr}({\mathbf{t}}^e)\cap \{0,1\}^{{\rm aou}(e,s)}]\res q+1=\{0,1\}^{q+1}$.
\end{proof}

Suppose for the sake of contradiction that $Z_1\circ Z_0\leq_{\rm W}{\rm AoUC}_{[0,1]}^k$ holds via computable $H$ and $K=\lrangle{K_0,K_1}$, i.e., given $(T,S)$, for any $k$-tuple $\mathbf{p}$ of infinite paths through trees $H(T,S)=\{H_i(T,S)\}_{i<k}$, $(K_0(\mathbf{p},T,S),K_1(\mathbf{p},T,S))\in Z_1\circ Z_0(T,S)$, that is, $K_0(\mathbf{p},T,S)\in[T]$ and $K_1(\mathbf{p},T,S)\in S(K_0(\mathbf{p},T,S))$.
Choose a computable function $f$ such that $\mathbf{t}^{f(e)}=H(T_e,S_e)$, $\varphi_{f(e)}=\lambda\mathbf{p}.K_0(\mathbf{p},T_e,S_e)$, and $\psi_{f(e)}=\lambda\mathbf{p}.K_1(\mathbf{p},T_e,S_e)$.
By Kleene's recursion theorem, there is $r$ such that $(\mathbf{t}^{f(r)},\varphi_{f(r)},\psi_{f(r)})=(\mathbf{t}^{r},\varphi_{r},\psi_{r})$.
This triple clearly satisfies the premise of the above claim.
The realizer $G$ in the claim witnesses the failure of $Z_1\circ Z_0\leq_{\rm W}{\rm AoUC}_{[0,1]}^k$ via $H$ and $K$.
Consequently, ${\sf LLPO}\star{\rm AoUC}_{[0,1]}\not\leq_{\rm W}{\rm AoUC}_{[0,1]}^k$ for all $k\in\mathbb{N}$.
\end{proof}
\end{theorem}

We point out that the preceding theorem relativizes, i.e.~even provides a separation w.r.t.~continuous Weihrauch reductions.

\begin{corollary}
\label{corr:maintheo}
$\llpo \star \aouc \nleqW \aouc^*$.
\begin{proof}
Assume to the contrary that $\C_{\{0,1\}} \star \aouc \leqW \aouc^*$. Consider as input to $\llpo \star \aouc$ the set $\uint$ together with the constant function $h : \uint \to \mathcal{A}(\{0,1\})$, $x \mapsto \{0,1\}$. The latter can be represented in such a way that it shares arbitrarily long prefixes with names for any other continuous function of that type. The reduction has to chose some $k \in \mathbb{N}$ eventually that serves as the first component of the derived input to $\aouc^*$. But since the original input can still be altered to any other suitable input, this would imply $\llpo \star \aouc \leqW \aouc^k$, thus contradicting Theorem \ref{theo:main}.
\end{proof}
\end{corollary}

\begin{corollary}
$\rDiv$ is not finitely concurrent.
\begin{proof}
By Proposition \ref{bimatrix:prop:aouequivrdiv}, $\rDiv \equivW \aouc$ and by Proposition \ref{prop:rdivprior} $\llpo \equivW \Ctwo \leqW \rDiv$. Thus, Corollary \ref{corr:maintheo} implies $\rDiv \star \rDiv \nleqW \rDiv^*$.
\end{proof}
\end{corollary}

We will find next that $\rDiv$ only barely fails being finitely concurrent: While some amount of nesting is required to obtain the full power of finitely many uses of $\rDiv$, nesting depths $2$ already suffices. This result will be proven via a number of individual technical contributions.

Let $\mathcal{O}(\mathbb{N})$ denote subsets of $\mathbb{N}$ represented via an enumeration of their elements. Call a set $A \subseteq \mathcal{O}(\mathbb{N})$ \emph{nice}, if $\emptyset \in A$ and $A$ contains a computable dense sequence $(a_n)_{n \in \mathbb{N}}$.

\begin{proposition}
\label{prop:cnextraction}
Let $f : \subseteq \mathcal{O}(\mathbb{N}) \mto \mathbf{X}$ have a nice domain and a computable closed graph. Then $f \star \aouc^k \leqW \C_{\{1,\ldots,2^k\}} \star \left (f^{2^k} \times \aouc^k \right )$.
\begin{proof}
For any subset $I \subseteq \{1,\ldots,k\}$ we compute an input to $f$ under the assumption that the components $i \in I$ for $\aouc^k$ are singletons, and the components $i \notin I$ are the whole interval. We start with providing a name for $\emptyset \in \dom(f)$ and wait until all components $i \in I$ have started to collapse. Then we can compute the actual values in those singletons, and can attempt to compute the input to $f$ associated with those values, together with $0 \in \uint$ for those components $i \notin I$. Before actually fixing any values, we make sure that there is some element $a_n$ of the dense sequence extending the current finite prefix. If we ever find that some component $i \notin I$ is starting to collapse, we abandon the attempt to find the correct input to $f$, and just extend the current prefix to some suitable $a_n$. By the assumption that $\dom(f)$ is nice, this is guaranteed to produce a valid input to $f$, and if $I$ was indeed the correct choice, will be the correct input.

Now we consider the output of $f$ on each of these values, together with the output($x_1,\ldots,x_k)$ of $\aouc^k$ on the original input. We replace those $x_i$ with $i \notin I$ with $0$, and ask whether this is still a correct output. As the graph of $f$ is a computable closed set, we can ask whether this output matches the input to $f$ obtained from the so modified output of $\aouc^k$. We can compute a truth value $t_I \in \mathbb{S}$ which is false iff both questions answer to true. If $I$ was indeed correct, the corresponding $t_I$ will be false. If $t_I$ is false, then the combined outputs of $f$ and $\aouc^k$ allow us to solve the original question to $f \star \aouc^k$. We can use $\C_{\{1,\ldots,2^k\}}$ to pick some false $t_I$.
\end{proof}
\end{proposition}

\begin{corollary}
\label{corr:cnextraction}
$\aouc^l \star \aouc^m \leqW \C_{\{1,\ldots,2^m\}} \star \aouc^{l2^m + m}$.
\begin{proof}
We just need to argue that $\aouc$ is equivalent to some $f : \subseteq \mathcal{O}(\mathbb{N}) \mto \mathbf{X}$ satisfying the criteria of Proposition \ref{prop:cnextraction}. Recall that $A \in \mathcal{A}(\uint) \supseteq \dom(\aouc)$ is represented by enumerating rational open balls exhausting the complement of $A$. By letting $f$ be equal to $\aouc$, but acting on the enumerations rather than the sets themselves, we have found the required candidate.
\end{proof}
\end{corollary}

\begin{proposition}
\label{prop:cnabsorbtion}
Let $f :\subseteq \mathcal{O}(\mathbb{N}) \mto \mathbf{X}$ have a nice domain. Then $f \star \C_{\{1,\ldots,n\}} \leqW f^n \times \C_{\{1,\ldots,n\}}$.
\begin{proof}
For each $i \in \{1,\ldots,n\}$ we attempt to compute the suitable input to $f$ if $i$ were the output provided by $\C_{\{1,\ldots,n\}}$. We only actually write a finite prefix of the output once we have found an element $a_n$ extending it. If we ever learn that $i$ is not a correct output of $\C_{\{1,\ldots,n\}}$, we abandon the attempt and simply extend the current input to $f$ to some $a_n$. The nice domain of $f$ ensures that this procedure results in a valid input for $f$. If we do this for all choices of $i$ in parallel, and also compute a suitable $i$, we can then read of a correct output to $f \star \C_{\{1,\ldots,n\}}$.
\end{proof}
\end{proposition}

\begin{corollary}
\label{corr:cnabsorbtion}
$\aouc^l \star \C_{\{1,\ldots,m\}} \leqW \aouc^{lm + m - 1}$.
\begin{proof}
To argue that we may use $\aouc$ in place of $f$ in Proposition \ref{prop:cnabsorbtion}, we argue as we did to obtain Corollary \ref{corr:cnextraction} from Proposition \ref{prop:cnextraction}. Now $\C_{\{1,\ldots,m\}} \leqW \Ctwo^{m-1}$ and $\Ctwo \leqW \aouc$ from \cite{paulyincomputabilitynashequilibria} complete the argument.
\end{proof}
\end{corollary}

So we do find that 3 (or more) consecutive applications of powers of $\aouc$ do reduce to $2$:

\begin{corollary}
\label{corr:aoucabsorbtion}
$\aouc^l \star \aouc^m \star \aouc^k \leqW \aouc^{(l + 1)2^k - 1} \star \aouc^{m2^k + k}$
\end{corollary}

\begin{proposition}
\label{prop:functionnbounded}
Let $f : \mathbf{X} \mto \mathbb{N}$ be such that $n \in f(x) \wedge m > n \Rightarrow m \in f(x)$. Then if $f \leqW \C_\Cantor$, $f$ is already computable.
\longversion{\begin{proof}
As $\C_\Cantor$ is a cylinder, we even have $f \leq_{\textrm{sW}} \C_\Cantor$. Thus, for any $x \in \mathbf{X}$ we obtain some $A_x \in \mathcal{A}(\Cantor)$ and some $K_x : A_x \to \mathbb{N}$ such that $K_x(p) \in f(x)$ for any $p \in A$. As $\Cantor$ is computably compact, so is $A_x$ in a uniform way, and we can thus compute $\left ( \sup_{p \in A_x} K_x(p) \right ) \in \mathbb{N}_>$. As from any $n \in \mathbb{N}_>$ we can compute some $m \in \mathbb{N}$ with $m \geq n$, the assumptions on $f$ thus imply computability of $f$.
\end{proof}}
\end{proposition}

\begin{corollary}
\begin{align*}
& \aouc^* \star \aouc^* & \equivW & \coprod_{n \in \mathbb{N}} (\aouc^n \star \aouc^n) \\ \equivW & \coprod_{n \in \mathbb{N}} \aouc^{(n)} \equivW \C_{\{0,1\}}^* \star \aouc^* & \equivW & \left (\aouc^* \right )^{(n+1)}\end{align*}
\begin{proof}
By Proposition \ref{prop:functionnbounded} it follows that in e.g.~$\aouc^* \star \aouc^*$ the number of oracle calls made in the second round can be bounded in advance. The equivalences now follow from the uniform versions of Corollaries \ref{corr:cnextraction}, \ref{corr:aoucabsorbtion}.
\end{proof}
\end{corollary}

\section{Gaussian Elimination}
Most work on algorithms in linear algebra assumes equality to be decidable, and is thus applicable to computability over the rational or algebraic numbers, but not to computability over the real numbers. In the latter setting, computability of some basic questions (rank, eigenvectors,\ldots) was studied in \cite{ziegler}, with some additional results in \cite{ziegler5,paulybrattka2}. Here, we shall consider \textrm{LU}-decomposition and Gaussian elimination.

Gaussian elimination is one of the basic algorithms in linear algebra, used in particular to compute the \textrm{LU}-decomposition of matrices. The goal is to transform a given matrix into row echelon form by means of swapping rows (and maybe columns) and adding multiples of one row to another. Sometimes the leading non-zero coefficients in each row are required to be $1$, however, as this is easily seen to require equality testing, we shall not include this requirement.

\begin{definition}
$\textrm{LU-Decomp}_{P,Q}$ takes as input a matrix $A$, and outputs permutation matrices $P$, $Q$, a matrix $U$ in upper echelon form and a matrix $L$ in lower echelon form with all diagonal elements being $1$ such that $PAQ = LU$. By $\textrm{LU-Decomp}_{Q}$ we denote the extension where $P$ is required to be the identity matrix.
\end{definition}

\begin{theorem}
$\textrm{LU-Decomp}_{P,Q} \equivW \rDiv^*$ and $\rDiv^* \leqW \textrm{LU-Decomp}_{Q} \leqW \rDiv^* \star \rDiv^*$.
\end{theorem}

The proof of the preceding theorem follows in form of some lemmata. We point out that the upper bounds are proven via variants of Gaussian elimination. In the case of $\textrm{LU-Decomp}_{P,Q}$ and its matching lower bound, this shows that Gaussian elimination exhibits no more incomputability than inherent in the problem it solves. It is consistent with the classifications that the extra freedom in choosing the pivot elements in solving $\textrm{LU-Decomp}_{P,Q}$ compared to solving $\textrm{LU-Decomp}_{Q}$ makes the problem less incomputable. Resolving the precise degree of $\textrm{LU-Decomp}_{Q}$ seems to be beyond the reach of our current methods though.

\begin{lemma}
$\textrm{LU-Decomp}_{P,Q} \equivW \textrm{LU-Decomp}_{P,Q}^*$ and $\textrm{LU-Decomp}_{Q} \equivW \textrm{LU-Decomp}_{Q}^*$.
\begin{proof}
An \textrm{LU}-decomposition of $\left (\begin{array}{cc} A & 0 \\ 0 & B\end{array} \right )$ gives rise to \textrm{LU}-decompositions of both $A$ and $B$.
\end{proof}
\end{lemma}

\begin{lemma}
$\textrm{LU-Decomp}_{P,Q} \leqW \rDiv^*$.
\begin{proof}
Initially, we rearrange all the matrix elements such that at each step the pivot element chosen has the largest absolute value amongst the remaining elements, and obtain their signs. This can be achieved by $\C_{\{0,\ldots,k\}}$ for suitable $k$. Then we can compute all the relevant divisions simultaneously, using some $\rDiv^l$. Corollary \ref{corr:cnabsorbtion} then shows that this reduces to $\rDiv^*$.
\end{proof}
\end{lemma}

\begin{lemma}
$\textrm{LU-Decomp}_{Q} \leqW \rDiv^* \star \rDiv^*$.
\begin{proof}
Given some real matrix $(a_{ij})_{i\leq n, j \leq m}$ we can use $\C_{\{0,\ldots,n-1\}}$ to pick some $i_0$ such that $|a_{i_0,1}| = \max_{i \leq n} |a_{i,1}|$, and permute the rows to move the $i_0$-th row to the top. We can use $\Ctwo^n$ to figure out for each $i$ whether $|a_{i,1}|$ is non-negative or non-positive. For each $i \neq i_0$ we compute $\rDiv(|a_{i,1}|,|a_{i_0,1}|)$, pick the sign depending on the putative signs on $a_{i,1}$ and $a_{i_0,1}$ and then subtract the corresponding multiple of the $i_0$-th row from the $i$-th row. By choice of $i_0$, either all $a_{i,1}$ are $0$ anyway, or $a_{i_0,1} \neq 0$ -- in both cases, this ensures that in all rows but the $i_0$-th the first entry is zero after the operation.

The procedure so far made use of $\rDiv^{n-1} \star \left ( \C_{\{0,\ldots,n-1\}} \times \Ctwo^n \right )$.

After the first round, the now first row is fixed. Amongst the remaining ones, we pick one with the largest absolute value in the second column (using $\C_{\{0,\ldots,n-2\}}$), determine the signs of entries in the second column (using $\Ctwo^{n-2}$) and again use $\rDiv$ to compute the coefficients for subtracting the second row from the lower ones.

This is repeated until each row has been dealt with. Overall, we use $n-1$ rounds, so the procedure is reducible to $\left [\rDiv^{n-1} \star \left ( \C_{\{0,\ldots,n-1\}} \times \Ctwo^n \right )\right]^{(n)}$. By repeated application of Corollaries \ref{corr:cnabsorbtion},\ref{corr:aoucabsorbtion} this reduces to $\aouc^k \star \aouc^k$ for sufficiently big $k$ (depending effectively on $n$).
\end{proof}
\end{lemma}

\hide{
\begin{proposition}
$\Ctwo^* \leqW \gauss$.
\begin{proof}
Given $\varepsilon_1,\ldots,\varepsilon_n,\delta_1,\ldots,\delta_n \in \mathbb{R}$, we compute the $2n \times n$ matrix:
\[\begin{bmatrix}
    \varepsilon_1 & 0 & 0 & \dots  & 0 \\
    \delta_1 & 0 & 0 & \dots  & 0 \\
    0 & \varepsilon_2 & 0 & \dots & \vdots \\
    0 & \delta_2 & 0 & \dots & \vdots \\
    \vdots & \vdots & \vdots & \ddots & \varepsilon_n \\
    0 & 0 & 0 & \dots  & \delta_n
\end{bmatrix}\]

Applying $\gauss$ to this matrix boils down to choosing for each $i \leq n$ one of $\varepsilon_i$ and $\delta_i$ as the leading entry of the corresponding row, and reducing the other to $0$ by a suitable subtraction. By inspecting the row-swapping operations, we can detect which one was chosen. Moreover, if one of $\delta_i$ and $\varepsilon_i$ is not zero, then the chosen one must be not zero, too. Thus, the non-chosen one is zero if there is a zero value. This in turn is just $\llpo(\varepsilon_i, \delta_i)$, and $\llpo$ is equivalent to $\Ctwo$.
\end{proof}
\end{proposition}}

\begin{lemma}
$\rDiv \leqW \textrm{LU-Decomp}_{P,Q}$.
\begin{proof}
We consider the computable function $B : \uint \to \mathbb{R}^{2 \times 2}$ of matrices defined via:
\[B(\varepsilon) = \exp(-\varepsilon^{-2}) \left (\begin{array}{cc} \cos(\varepsilon^{-1}) & \sin (\varepsilon^{-1}) \\ - \sin (\varepsilon^{-1}) & \cos(\varepsilon^{-1}) \end{array} \right ) \textnormal{ for } \varepsilon > 0 \quad B(0) = \left (\begin{array}{cc} 0 & 0 \\ 0 & 0 \end{array} \right )\]
This is based on a counterexample due to \name{Rellich} (cmp.~\cite[II.5.3]{kato}, \cite[Example 18]{ziegler}). If $\varepsilon \neq 0$, then the lower-left corner of $L$ in an \textrm{LU}-decomposition of $B(\varepsilon)$ will be one of $\tan \varepsilon^{-1}$, $\cot \varepsilon^{-1}$ or $-\cot \varepsilon^{-1}$. The relevant case can be obtained from $P$ and $Q$. As $\arctan$ and $\arccot$ are total, we can apply the relevant inverse even if $\varepsilon = 0$, and thus the lower-left corner of $L$ is an arbitrary real number. Let $x'_\varepsilon$ be the result, and $x_\varepsilon = \max \{0, \min \{1, x'_\varepsilon\}\}$.

We want to show that $\aouc \leqW \gauss$ (which is equivalent to the claim by Proposition \ref{bimatrix:prop:aouequivrdiv}). Given $A \in \dom(\aouc)$, we show how to compute some $\varepsilon \in \uint$ such that $x_\varepsilon \in A$. As long as $A = \uint$ is consistent with our knowledge of the input, we specify that $\varepsilon \in [0,2^{-t}]$ for smaller and smaller $t \in \mathbb{N}$. If we learn at time $t$ that $A \neq \uint$, we compute $y$ such that $A = \{y\}$ and choose $k \in \mathbb{N}$ such that $(2k\pi)^{-k} \leq 2^{-t}$. We can then specify $\varepsilon = (2k\pi + y)^{-1}$. But now $x_\varepsilon = y$. If $A = \uint$, then $x_\varepsilon \in A$ anyway by definition.
\end{proof}
\end{lemma}


\bibliographystyle{eptcs}
\bibliography{../spieltheorie}


\end{document}